\documentclass[12pt]{article}
\begin{document}
\title{A new proof for non-occurrence of trapped surfaces and  information paradox}
\author{Abhas Mitra\\ Nuclear Research Laboratory, Bhabha Atomic
Research Centre\\ Mumbai-400085, India\\ E-mail: amitra@apsara.barc.ernet.in,
abhasmitra@rediffmail.com}
\date{}
\maketitle
\begin{abstract}

We present here a very simple, short and new 
proof which shows that no {\em trapped surface}
is ever formed in spherical gravitational collapse of isolated bodies.
Although this derivation is of purely mathematical nature and without any assumption,
 it is shown, in the Appendix, that,
physically, trapped surfaces do not form in order that the 3 speed of the fluid
as measured by an observer at a fixed circumference coordinate $R$ (a scalar),
is less than the speed of light $c$. The consequence of this result is that,
mathematically, even if there would be Schwarzschild Black Holes,
they would have unique gravitational mass $M=0$. Recall that Schwarzschild BHs
may be considered as a special case of rotating Kerr BHs with rotation
parameter $a=0$. If one would derive the Boyer-Lindquist metric [1]
in a straight forward manner
by using the Backlund transformation[2], one would obtain $a= M \sin \phi$ where $\phi$
is the azimuth angle. This relation demands that BHs have unique mass
$M=0$ (along with $a=0$) which in turn confirms that there cannot be
any trapped surface in realistic gravitational collapse where the fluid has
real pressure and density. Since there is no trapped surface and horizon,
there is no Information Paradox in the first place.
\end{abstract}
When a  self-gravitating fluid undergoes gravitational contraction, by virtue
of Virial Theorem,
part of the gravitational energy 
must be radiated out. Thus  the total mass energy, $M$,
($c=1$) of a body decreases as its  radius $R$ decreases.  But in Newtonian regime ($2M/R\ll 1$, $G=1$),
$M$ is almost fixed and  the evolution of the 
ratio, $2 M/R $,  is practically dictated entirely by $R$. 
 If it is {\em assumed} that even in the extreme general relativistic
case $2M/R $ would behave in the {\em same Newtonian} manner,  then for sufficiently
small $R$, it would be
possible to have  $2M/R > 1$, i.e, trapped
surfaces would form[3,4]. 

Unfortunately, even when we use General Relativity (GR), our intuition is often governed
by Newtonian concepts, and thus,
intuitively, it appears that, as a fluid would collapse, its gravitational mass would
remain more or less constant so that for continued collapse, sooner or later, one
would have $2 M/R >1$, i.e, a ``trapped surface'' must form. The singularity
theorems thus start with the {\em assumption} of formation of trapped surfaces[3,4].
However, many readers (from experimental astronomy, particle physics,
quantum-gravity and quauntum information fields) may not be even aware
that {\em occurrence of trapped surface is a conjecture} and the Singularity Theorems
are based on this conjecture. Further, in the literature, there have been attempts
to find out ``sufficient and necessary condition'' for realization of this conjecture
of trapped surfaces for spherical collapse[3] without showing that such ``necessary and
sufficient conditions'' {\em are fullfilled}. Nonetheless, we have found that, by superficially
going through such papers, many readers (erroneously) think that the conjecture
of trapped surface has been proved. The actual situation regarding this has been succinctly
brought out in the following statement[5]:

``it is necessary to point out that a demonstration of the trapped surface conjecture remains
elusive.''

In fact, there are specific examples
that trapped surfaces do not form. For example, in the cosmological context,
Nariai Metric[6] is the first specific
example of  non-occurrence of trapped surfaces.  In this context, 
Dadhich's comment
 about this
``assumption'' of trapped surfaces is[7]:

``It is noteworthy that violation of this assumption entailed no unphysical
features. This assumtion seriously compromises, as is demonstrated by this
example, the generality of the theorems (i.e, singularity theorems, author).
It has always been looked upon as suspect.''

Another work on spherical collapse using premeditated  specific metric finds
that collapse would not produce any Horizon because of heat flow, i.e, because
of decrease of $M$ during collapse[8]. This specific example focussed attention only
at the boundary of the fluid and had it treated the inner mass shells, it might
have found absence of trapped surfaces too.

In 1990, Senovilla constructed a specific model of a cylindrically symmetric universe
without any trapped surface[9]. In 2002, Goncalves, in a more general manner, showed
the absence of trapped surface and singularities in cylindrical collapse[10].

In the cosmological context, $M$ remains fixed (because radiation cannot leave the universe)
while for isolated bodies, $M$ necessarily decreases due to emission of radiation,
and thus, it is more likely that trapped surfaces do not form.

In any case, once  the formation of a ``trapped surface'' is assumed, then, a set of very 
 reasonable assumptions would show that the collapse would become irreversible. 
Physically,
once trapped surface formation is {\em assumed}, it would appear that the sign of the
pressure gradient force would reverse and thus pressure would aid the collapse
rather than hinder the same.
 Furthermore, since outgoing radiation too would turn inward,
 no mass energy would escape 
and  the 
mass energy enclosed within a shell $M(r)$, where $r$ is a comoving coordinate, 
 would not decrease any longer. Hence if the collapse would lead to a BH with an all emcompassing
Event Horizon (EH), it would naturally appear that the mass of the BHs must be finite. Thus
the intuitive and {\em apparently} correct notion of finite mass BHs rests on the {\em assumption}
of formation of trapped surfaces, which, in turn, rests on our intuitive
(but incorrect)
idea that even when  gravity would be so strong as to  trap even light $M$ would remain more or less
unchanged 
{\em as in the Newtonian case}. However, the following exact treatment would unequivocally
show that, trapped surfaces do not form in any spherical collapse.

If we take the signature of spacetime as (+,-,-,-)
the spherically symmertical metric for an arbitrary fluid, in terms of comoving
coordinates $t$ and $r$ is[3,4,12]
\begin{equation}
ds^2 = g_{00} dt^2 + g_{rr} dr^2 - R^2 (d\theta^2 + \sin^2\theta d\phi^2)
\end{equation}
where $R=R(r, t)$ is the Invariant Circumference coordinate and happens to be
 a scalar. Further,  for radial motion with $d\theta =d\phi =0$,
the metric becomes

\begin{equation}
ds^2 = g_{00} dt^2 (1- x^2)
\end{equation}
or,

\begin{equation}
(1-x^2) = {1\over g_{00}} {ds^2\over dt^2}
\end{equation}
where the auxiliary parameter

 \begin{equation}
 x = {\sqrt {-g_{rr}} ~dr\over \sqrt{g_{00}}~ dt}
 \end{equation}
For a fluid element at $r=fixed$, obviously $x=0$ because $dr=0$.
However,  $dr$, in general, is not zero; otherwise the metric in comoving coordinates
would be
\begin{equation}
ds^2 = g_{00} dt^2  - R^2 (d\theta^2 + \sin^2\theta d\phi^2)
\end{equation}
and which is {\em not} the case.
 The comoving observer at $r=r$ is
free to do measurements of not only the fluid element at $r=r$ but also of other objects:
  If the
comoving observer is compared with a static floating boat in a flowing
river, the boat can monitor the motion of other boats or the pebbles fixed
on the river bed. Here the fixed markers on the river bed are like the
background $R=  constant$ markers against which the
river flows. If we intend to find the parameter $x$ for such a $R=constant$
marker, i.e, for a pebble lying on the river bed at a a {\em fixed} $R$, we
will have
 
 \begin{equation}
 d R(r,t) = 0= {\dot R} dt + R^\prime dr 
 \end{equation}

where an overdot denotes a partial derivative w.r.t. $t$ and a prime denotes
a partial derivative w.r.t. $r$.
Therefore for the $R=constant$ marker, we find that
\begin{equation}
{dr\over dt} = - {{\dot R}\over R^\prime}
\end{equation}
and the corresponding $x=x_c$ is

\begin{equation}
x= x_{c} = {\sqrt {-g_{rr}} ~dr\over \sqrt{g_{00}}~ dt} = -{\sqrt {-g_{rr}}
~{\dot R}\over \sqrt{g_{00}}~ R^\prime}
 \end{equation}
Using Eq.(3), we also have
\begin{equation}
(1-x_c^2) = {1\over g_{00}} {ds^2\over dt^2}
\end{equation}
Now let us define[4]
\begin{equation}
\Gamma = {R^\prime\over \sqrt {-g_{rr}}}
\end{equation}
and[4]
\begin{equation}
U = {{\dot R}\over \sqrt{g_{00}}}
\end{equation}

so that the combined Eqs. (8), (10) and (11) yield
\begin{equation}
x_c = {U\over \Gamma}; \qquad U= -x_c \Gamma
\end{equation}
As is well known, the gravitational mass of the collapsing (or expanding) fluid is defined
through the equation[4,12]
\begin{equation}
\Gamma^2 = 1 + U^2 - {2M(r,t)\over R}
\end{equation}
Then the two foregoing equations can be combined and transposed to obtain

\begin{equation}
\Gamma^2 (1- x_c^2) = 1- {2M(r,t)\over R}
\end{equation}

By using Eqs.(9) and (10) in the foregoing Eq., we have

\begin{equation}
{{R^\prime}^2\over {-g_{rr} g_{00}}} {ds^2\over dt^2} = 1 - {2M(r,t)\over R}
\end{equation}

Recall that the determinant of the metric tensor is always negative[11]:
\begin{equation}
g = R^4 \sin^2 \theta ~g_{00} ~g_{rr} \le 0
\end{equation}
so that we must always have
\begin{equation}
-g_{rr}~ g_{00} \ge 0
\end{equation}

Further for the signature chosen here, $ds^2 \ge 0$ for all material
particles or photons. Then noting Eq.(17), it follows that the LHS of  Eq. (15) is {\em always positive}.
So must then be the RHS of the same Eq. which implies that

\begin{equation}
{2M(r,t)\over R} \le 1
\end{equation}

This shows in the  utmost general fashion that trapped surfaces
are not formed in spherical collapse or expansion. As such this result
 is already known[12,13,14]. However, several readers have
found the papers[12,13] too long and the central derivation to be rather
complicated. Further when we wrote the papers [12,14], we failed to bring out
the precise physical implication of the parameter ''${\bf v}$'' appearing there.
We have removed this shortcoming in the present derivation. The Appendix I. 
will show, in a very transparent manner, that
 ${\bf v}$ is the 3-speed of the fluid with respect to an observer sitting
at a fixed $R$. Nevertheless, it may be noted that to arrive at the no trapped surface
condition, we do not require this physical interpretation of ${\bf v}$.

In the absence of any trapped surface, a collapsing fluid will always keep on radiating and $M(r)$ would keep
on decreasing. In case it would be {\em assumed} that, the collapse would continue
all the way upto $R=0$, then the constraint (18) demands that $M(R=0) =0$ too.
And this is the reason that all BHs (even if they would
be assumed to exist) must have $M=0$.

  Once there would be no trapped surface, virial theorem would ensure
that the collapse process causes not only emission of radiation but there is automatic
increase of internal energy and attendant pressure gradient. Thus even if there may not
be any stable equilibrium, there could be approximate quasistable states which could be
supported by the collapse generated internal pressure gradient. Eventhough the gravitational
mass would go on decreasing, the field strength would ever increase. This would
keep on enlarging the proper radial depth (stretching of spacetime membrane) and
collapse would continue indefinitely in the inner (proper) spacetime eventhough, externally,
the mouth of the potential well $R \to 0$[12,13,15, 16]. The equality of Eq.(18) can be
satisfied only at this final stage, i.e, an apparent horizon can form as
$R\to R_H \to 0$ and $M \to 0$[12,13,15,16]. This picture is corroborated by
a recent work which shows that for spherical gravitational collapse, it is possible to
have a situation where[5]:

`` the proper distance in the transformed metric from points `near' the horizon
to the horizon itself becomes infinite. However, the area of the spheres ($4 \pi R^2$,
$R\to R_H \to 0$, author) does not change ($R$ hovering around $0$) because the angular
metric components are unaffected. This means that the manifold `near' the horizon gets
transformed into an infinitely long cylinder (inner spacetime, author) whose
crosssection asymptotes to the original area of the horizon and the three-scalar-curvature
along the cylinder is a positive constant ($\sim 1/R_H^2$)..''

As $R\to R_H \to 0$, the scalar curvature blows up. And as the proper radial depth
$l \to \infty$, the local observer's proper time of collapse $\tau \to \infty$ and
 the
collapse becomes Eternal.

However in a finite proper time, there would always be a finite $M$ and finite $R$ 
associated with an Eternally Collapsing Object (ECO). It is quite likely that the so-called
Black Hole Candidates are actually ECOs. Since the ECOs are expected to be

(i) Hot, i.e, supported largely by either trapped radiation (photons and neutrinos) pressure
or freshly generated radiation pressure due to contraction at unimaginable slow rate and

(ii) Not in strict hydrostatic equilibrium, i.e, they may be collapsing in a strict sense
in the same way as primordial clouds or pre-main sequence stars are collapsing

the conventional mass upper limit ($M_{max} \sim 3-4 M_\odot$) of {\em cold} objects
is not applicable to them. On the intergalactic scale, the BHCs could simply be
something like {\em hot} Suppermassive Stars, which, however, are contracting 
at incredibly slow rate
and generating radiation pressure even in the absence of nuclear fuel [12,13,14,15,16].

All astrophysical plasma is neutral on a macroscopic scale and thus even if one would
imagine existence of finite mass BHs, they would be chargeless and without any
intrinsic magnetic field. On the other hance since even neutral astrophysical
plasma is expected to generate intrinsic magnetic field, in the absence of any EH, the
ECOs are expected to have strong intrinsic magnetic field. This prediction that the
BHCs should have strong intrinsic magnetic field has also been verified[17,18,19].

And when there is no BH or no EH, there is never any trapping of Quantum Information
let alone vanishing of the same from the universe[11, 12]. Thus the entire debate
on the supposed Quantum Information Paradox and its supposed resolution is
meaningless. Since there is no singularity (in a finite proper time), neither
will there be any ``White Hole'', ``Baby Universe'', ``Worm Hole'', ``Time Machine''
or any other science fiction item.

Finally recall that, the Schwarzschild BHs can be considered as a special case
of rotating Kerr BHs with the rotation parameter $a=0$. And if one would derive the
Boyer-Lindquist metric for rotating BHs[1] in a straight forward way by using the
Backlund transformation[2], it would be found that $a$ and $M$ are interlinked through

\begin{equation}
a = M \sin\phi
\end{equation}

Since $a$ and $M$ are constants while $\phi$ is not so, in order to satisfy this relationship,
we must have

\begin{equation}
a=0
\end{equation}

as well as

\begin{equation}
M=0
\end{equation}

The last equation demands that there is no trapped surface as already shown. Any
reader not having access to the ref.[2] containing Eq.(19) is welcome to contact the author
for a scanned image of the same.

\subsection{ Appendix I: Physical significance of $x$ and $x_c$}
Although it is not necessary, yet, it would be worthwhile to understand the physical reason for non-occurrence
of trapped surfaces. To this effect, 
first note the most
 general form of a metric[11]
\begin{equation}
ds^2 = g_{00} (dx^0)^2 + g_{\alpha \beta} dx^{\alpha} dx^\beta + 2 g_{0 \alpha} dx^0 dx^\alpha
\end{equation}
Here $\alpha, \beta = 1,2,3$ represent the 3 spatial coordinates and $0$
represents the temporal coordinate.  Note that this general metric(22) can be seperared into a spatial and temporal part[11]:

\begin{equation}
ds^2 = d\tau_s^2 - dl^2
\end{equation}
where 
\begin{equation}
d\tau_s^2 = {g_{00}} (dx^0 - g_\alpha dx^\alpha)^2; \qquad g_\alpha = - {g_{0\alpha}\over g_{00}}
\end{equation}

and 
\begin{equation}
dl^2 = \left( -g_{\alpha \beta} + {g_{0 \alpha} g_{0 \beta} \over g_{00}}\right) dx^\alpha dx^\beta
\end{equation}
Here $d\tau_s$ is the element of {\em synchronized} proper time and $dl$ is
element of proper distance[11]. The arbitrary metric can also be rewritten as

\begin{equation}
ds^2 = d\tau_s^2 (1 -v^2)
\end{equation}

whence the 3-speed $v$ gets defined as[11]

\begin{equation}
v^2 = {dl^2\over d\tau_s^2}
\end{equation}

If for a specific case, such as a static metric or the present spherical case (which is
a non-stationary metric) 
 $g_{0 \alpha} =0$, we will have
\begin{equation}
d\tau_s^2 = d\tau^2 =  {g_{00}} ~dt^2 
\end{equation}

and
\begin{equation}
dl^2 =  -g_{\alpha \beta}  dx^\alpha dx^\beta
\end{equation}
where
 $d\tau $ is the usual proper time interval.
 Further, when all cross terms are zero (as in the present case), i.e, when $g_{\alpha \beta}$ too 
is diagonal 
\begin{equation}
dl^2 = -g_{rr}~ dr^2
\end{equation}
Then
\begin{equation}
v = {\sqrt {-g_{rr}}~ dr\over \sqrt{g_{00}} ~dt} 
 \end{equation}
Now going back to Eqs.(2) and (3), we quickly indentify $x$ as the 3-speed of an object (not necessarily fluid element)
measured by the comoving observer at $r=r$. Obviously, the 3 speed of the fluid element
itself, at $r=r$ is $x=0$.  But here $x_c$ is  the 3-speed of the $R=constant$ marker,
i.e, the pebble fixed on the river bed, and is non-zero.

Since $x_c$ is the speed of the $R=constant$ marker w.r.t. the $r=constant$ observer,
  the speed of the $r=constant$ observer, i,e, {\em  fluid itself},  w.r.t. the $R=constant$ marker
is the {\em negative} of $x_c$:
\begin{equation}
v = - x _{c} = -{\sqrt {-g_{rr}}~ dr\over \sqrt{g_{00}} ~dt} = +{\sqrt {-g_{rr}}
~{\dot R}\over \sqrt{g_{00}} ~R^\prime}
 \end{equation}
In terms of $v =  -x_c$, let us rewrite Eqs. (12) and (14) as

\begin{equation}
U = + v \Gamma
\end{equation}

and

\begin{equation}
\Gamma^2 (1- v^2) = 1- {2M(r,t)\over R}
\end{equation}
Now using Eq.(18) in (34), we find that both sides of it are positive and hence
\begin{equation}
v^2 \le 1
\end{equation}
Thus, in retrospect, we see that  non-occurrence of trapped surface
is a direct consequence of this  {\em cornerstone of relativity}: $ v^2 \le 1$
for material particles and photons.
If $\gamma$ is the corresponding Lorentz factor, the master equation (34) of
general relativistic fluid motion acquires the appealing form:
 
\begin{equation}
{\Gamma^2 \over \gamma^2} = 1 - {2 M (r,t)\over R}
\end{equation}


\vskip 2cm
\begin{thebibliography}{99}
\bibitem [1]{1}  Boyer, R.H. and Lindquist, R.W. Maximal analytic extension of the
 Kerr metric, {\it J. Math. Phys.} {\em 8}, 265 (1967)
\bibitem[2]{2} Neugbauer, G. in {\it General Relativity} (eds. G.S. Hall
\& J.R. Pulham) (SUSSP, Edinburg and IOP, London, 1996) (see p. 73 ).

\bibitem[3]{3} Bizon, P., Malec, E. and O'Murchadha, N.,
Trapped Surfaces in Spherical Stars, {\it Phys. Rev. Lett.} {\em 61(10)}, 1147 (1988).

\bibitem[4]{4} Misner, C.W., Thorne, K.S.. \& Wheeler, J.A., {\em Gravitation},
(Freeman, San Fransisco, California 1973).

\bibitem[5]{5} Malec, E. \& O'Murchadha, N., The Jang equation,
apparent horizons, and the Penrose inequality (2004) (gr-qc/0408044).
\bibitem[6]{6} Nariai, H. {\it Sci. Rep. Tohoku Univ.} {\em 34}, 160 (1950).
\bibitem[7]{7} Dadhich, N., Nariai metric is the first example
of the singularity free model (2003) (gr-qc/0106023).
\bibitem[8]{8} Banerjee, A., Chatterjee, S. \& Dadhich, N.,
Spherical collapse with heat flow and without horizon {\it Mod.
Phys. Lett. A} {\em 17}, 2335 (2002) (gr-qc/ 0209035).

\bibitem[9]{9} Senovilla, J.M.M., New class of inhomogeneous
cosmological perfect fluid solutions without big-bang singularity,
{\it Phys. Rev. Lett.} {\em 64}, 2219 (1990).

\bibitem[10]{10} Goncalves, S.M., Absence of trapped surfaces and singularities in cylindrical collapse,
{\it Phys. Rev. D} {\em 65}, 4045G (2002).

\bibitem[11]{11} Landau, L.D. \& Lifshitz, E.M., {\it The Classical Theory
of Fields} (Pergamon Press, Oxford, 1962).

\bibitem[12]{12} Mitra, A., Non-occurrence of trapped surfaces and black holes in
spherical collapse, {\it Found. Phys. Lett.} {\em 13(6)}, 543 (2000)
(astro-ph/9910408)
\bibitem[13]{13} Mitra, A., On the final state of spherical gravitational collapse,
{\it Found. Phys. Lett.} {\em 15(5)}, 439 (2002)
(astro-ph/0207056)

\bibitem [14] {14}  Leiter, D. \& Robertson, S. Does principle of equivalence
prevent trapped surfaces 
from being formed in general relativistic collapse process
 {\it Foun. Phys. Lett.} {\em 16}, 143 (2003) (astro-ph/0111421).


\bibitem[15]{15} Mitra, A. On the nature of the compact condensations at
the centre of galaxies, {\it Bull. Astr. Soc. India} {\em 30}, 173 (2002)
(astro-ph/020526).
\bibitem[16]{16} Mitra, A. On the question of trapped surfaces and black holes (2001)
(astro-ph/0105532).





\bibitem [17]{17}  Robertson, S. \& Leiter, D. Evidence for intrinsic 
magnetic moment in black hole candidates, {\it Astrophys. J.} {\em 565}, 447 (2002) (astro-ph/0102381).



\bibitem[18]{18}  Robertson, S. \& Leiter, D. On the intrinsic magnetic moment
in black hole candidates {\it Astrophys. J.} {\em 569}, L203 (2003) (astro-ph/0310078).

\bibitem [19]{19}  Robertson, S. \& Leiter, D. On the origin of 
the radio/X-ray luminosity correlation in black hole candidates,
{\it Mon. Not. Roy. Astr. Soc.} {\em 350}, 1391 (2004) (astro-ph/0402445).



\end{thebibliography}
\end{document}